\documentclass[journal,transmag]{IEEEtran}
\usepackage{hyperref,color}
\usepackage{amsmath,graphicx}
\usepackage[utf8]{inputenc}
\usepackage[top=0.7in, left=0.65in]{geometry}
\usepackage{cases}
\usepackage{cases}
\usepackage{amssymb}
\usepackage{siunitx}
\usepackage{amsthm}
\usepackage{amssymb}
\usepackage{eso-pic}

\newcommand{\flux}{\boldsymbol{\Psi}}

\AddToShipoutPicture*{\footnotesize\sffamily\raisebox{0.8cm}{\hspace{1.65cm}\fbox{\parbox{\textwidth}{\copyright~2017 IEEE. Personal use of this material is permitted. Permission from IEEE must be obtained for all other uses, in any current or future media, including reprinting/republishing this material for advertising or promotional purposes, creating new collective works, for resale or redistribution to servers or lists, or reuse of any copyrighted component of this work in other works.}}}}

\begin{document}
	\title{Optimized Field/Circuit Coupling for the Simulation of Quenches in Superconducting Magnets}
	\author{
		\IEEEauthorblockN{
			I. Cortes Garcia\IEEEauthorrefmark{1}, 
			S. Schöps\IEEEauthorrefmark{1}, 
			M. Maciejewski\IEEEauthorrefmark{2,3},
			L. Bortot\IEEEauthorrefmark{2}, 
			M. Prioli\IEEEauthorrefmark{2}, 
			B. Auchmann\IEEEauthorrefmark{2,4}, 
			and A.P. Verweij\IEEEauthorrefmark{2}
		}
		\IEEEauthorblockA{\IEEEauthorrefmark{1}Technische Universität Darmstadt, Darmstadt, Germany}
		\IEEEauthorblockA{\IEEEauthorrefmark{2}CERN, Geneva, Switzerland}
		\IEEEauthorblockA{\IEEEauthorrefmark{3}Łódź University of Technology, Łódź, Poland}
		\IEEEauthorblockA{\IEEEauthorrefmark{4}Paul Scherrer Institut, Villigen, Switzerland}
	}

	\IEEEtitleabstractindextext{%
	\begin{abstract}
	In this paper, we propose an optimized field/circuit coupling approach for the simulation of magnetothermal transients in superconducting 
	magnets. The approach improves the convergence of the iterative 
	coupling scheme between a magnetothermal partial differential model and an electrical lumped-element circuit. Such a multi-physics, multi-rate 
	and multi-scale problem requires a consistent formulation 
	and a dedicated framework to tackle the challenging transient effects occurring at both circuit and magnet level during normal operation and in 
	case of faults. We derive an equivalent magnet model at the 
	circuit side for the linear and the non-linear settings and discuss the convergence of the overall scheme in the framework of optimized Schwarz 
	methods. The efficiency of the developed approach is 
	illustrated by a numerical example of an accelerator dipole magnet with accompanying protection system.
	\end{abstract}
	
	}

	\maketitle
	\thispagestyle{empty}
	\pagestyle{empty}

	\section{Introduction}
	\IEEEPARstart{S}{uperconducting} magnets produce high magnetic fields used in high-energy particle accelerators for bending particle beams. In 
	order to reach the superconducting state, 
	the magnets are operated at very low temperatures ($1.9$\,K). Since the heat capacity is low at cryogenic temperatures, the magnets are prone to 
	quench due to a local energy deposition 
	(coupling losses in the superconducting cable, beam losses, cryogenic malfunction, mechanical movement, etc.). A quench is a transition from the 
	superconducting to the normal conducting 
	state. As a consequence, the release of the magnetic energy as Ohmic losses may result in a catastrophic damage in the magnet and circuit.

	The simulation of quench initiation, propagation, and subsequent protective measures represents a challenge in terms of the number of coupled physical domains, their highly non\-li\-ne\-ar behavior, their geometric scales, and their vastly different time constants. Quench protection systems such as the quench heaters \cite{Dahlerup-Petersen_1999aa} and the coupling-loss induced quench system \cite{Ravaioli_2015aa} are affecting both magnet and circuit. Therefore, their mutual influence has to be studied carefully. To this end, field-circuit coupling is inevitable. It brings together a magneto-thermal problem based on partial differential equations and a circuit model of the magnet's powering circuit. In the paper, we are presenting an equivalent superconducting magnet model on the circuit side capturing the magnetic and thermal behavior needed to represent a quench in the superconducting magnet.
	\begin{figure}
		\centering
		\includegraphics[width=.3\textwidth]{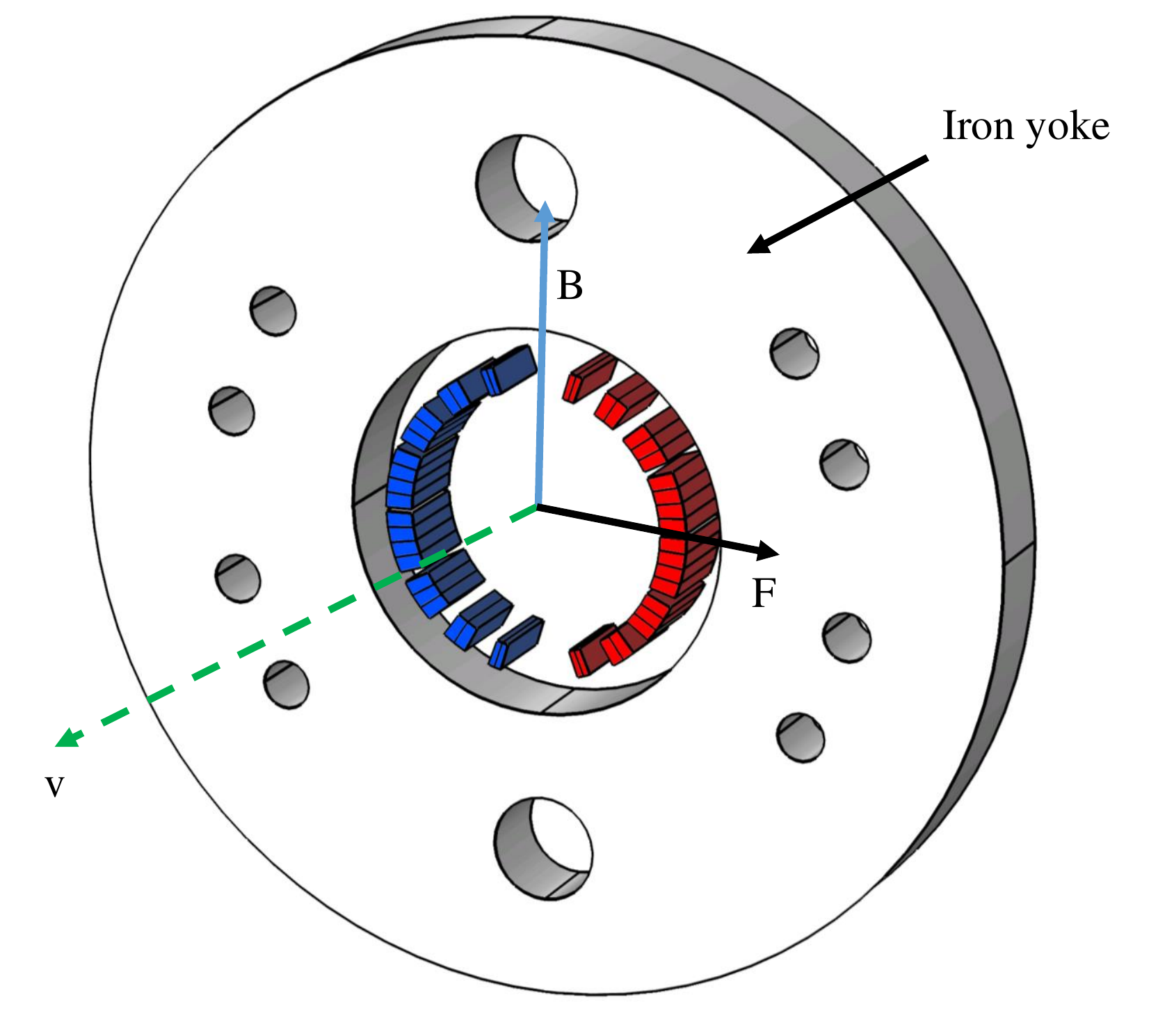}
		\vspace{-0.5em}
		\caption{Coil surrounded by an iron yoke. The current flowing through the coil 
		into the plane is depicted with red and its return path is marked
			with blue. Positively charged particle traveling in the middle of 
			the coil's aperture subject to vertical component of the magnetic field will be 
			deflected due to the Lorentz force (green dashed line).}
		\vspace{-1em}
		\label{fig:pics_multiscale_magnet}
	\end{figure}

	The coupling of electric fields and circuits is a well understood technique. Differences 
	among various approaches are how the coupled problem is 
	tackled. It is either solved `monolithically' as one system, such as e.g. 
	proposed in \cite{Dreher_1995aa} or it is coupled `weakly' using 
	additional 
	iterations or exploiting temporary linearizations 
	\cite{Bedrosian_1993aa,Lange_2009aa,Zhou_2006aa}. In most cases, the coupling is 
	established 
	via current sources, voltage sources or inductances. In \cite{Schops_2010aa,Bartel_2013aa} 
	the various approaches have been interpreted in the 
	context of waveform relaxation methods \cite{Lelarasmee_1982ab,Burrage_1995aa} such that 
	convergence could be proven.
	This paper interprets the coupling conditions in terms of optimized Schwarz methods, 
	where they are often called `transmission conditions', e.g. 
	\cite{Al-Khaleel_2014aa,Nshimiyimana_2016aa}. In contrast to prior works, 
	o numerical optimization is carried out, since it is cumbersome in the case of 
	multi-port devices and it is limited to linear cases. Instead, it is shown that 
	inductances are a first order approximation to the optimal condition. 
	
	The paper is structured as follows: Section 2 deals with the modeling of 
	superconducting magnets and their spatial discretization, Section 3 introduces 
	the idea of waveform relaxation and discusses the optimized contraction factor. 
	Section 4 underlines the findings by a real world simulation of an accelerator magnet 
	coupled to its surrounding protection circuitry. The paper closes with conclusions and 
	an outlook.

	\section{Modeling}
\begin{figure}
		\centering
		\includegraphics[width=0.6\linewidth]{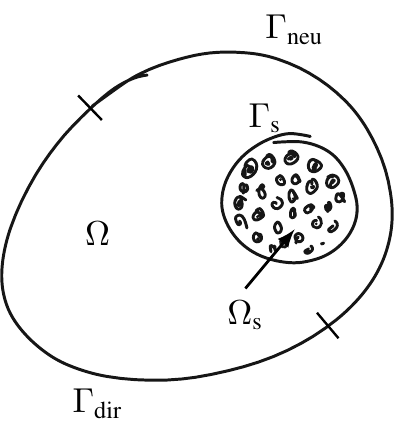}
		\caption{Sketch of the computational domain and subdomains with boundaries}
		\label{fig:patata}
	\end{figure}

	Starting from Maxwell's equations and assuming a magnetoquasistatic (MQS) setting, i.e., 
	neglecting displacement currents, the following partial differential equation 
	\begin{align}
	\!\nabla \times (\nu\nabla\times\vec{A})+\nabla\times\left(\nu_{\mathrm{s}}\tau_{\mathrm{eq}}\nabla\times\frac{\partial\vec{A}}{\partial 
		t}\right)
	= \vec{\chi}_{\mathrm{s}}\mathbf{i} \label{eq:mqs}
	\end{align} 
	is obtained on a domain $\Omega\in\mathbb{R}^{3}$ with an initial value $\vec{A}_0$ 
	at time $t_0$ and with Dirichlet and Neumann boundary 
	conditions on $\Gamma_\text{dir}\cup\Gamma_\text{neu}=\Gamma:=\partial\Omega$, 
	respectively. The eddy currents are taken into account in the coil 
	domain $\Omega_\text{s}$ by homogenization~\cite{Ravaioli_2015aa,Verweij_1995aa}, 
	see Fig.~\ref{fig:patata}. $\vec{A}$ is the magnetic
	vector potential, $\nu$ and $\nu_\text{s}$ are inverse magnetic permeabilities that 
	may depend on $\mathit{B} = \|\vec{B}\|$ with 
	$\vec{B}=\nabla\times\vec{A}$ the magnetic flux density and $\mathbf{i}$ the lumped 
	currents through each coil. In $\Omega_\text{s}$, the time 
	constant $\tau_{\mathrm{eq}}$ is defined such that the cable magnetization 
	$\vec{M}_{\mathrm{s}}$ satisfies the relation
	\begin{equation*}
	\nu_{\mathrm{s}}^{-1}\vec{M}_{\mathrm{s}} = \tau_{\mathrm{eq}}\frac{\partial\vec{B}}{\partial t},
	\end{equation*}
	where $\tau_{\mathrm{eq}}$ also depends on $B$. 
	Finally, $\vec{\chi}_{\mathrm{s}}:\Omega\mapsto\mathbb{R}^{3\times n_\text{s}}$, 
	with $n_\text{s}$ the number of coils, is the stranded-conductor 
	winding function (see \cite{Schops_2013aa}), which homogeneously distributes the 
	current in the domain $\Omega_\text{s}$ such that the current 
	density $\vec{J}_{\mathrm{s}} = \vec{\chi}_{\mathrm{s}}\mathbf{i}$. 
	The field equation \eqref{eq:mqs} is coupled to a circuit equation model via the 
	flux linkage
	\begin{equation}
	\mathbf{v} = \frac{\mathrm{d}}{\mathrm{d} t}\flux,
	\text{ with }
	\flux=\int_{\Omega}\vec{\chi}_{\mathrm{s}}^\top \vec{A} \ \mathrm{d} \Omega.\label{eq:mqs_coup}
	\end{equation}

	The temperature in the domain $\Omega_\text{s}$ can be obtained from the heat balance equation
	\begin{equation}
	\rho C_{\mathrm{p}}\frac{\partial T}{\partial t} - \nabla \cdot (k \nabla T) = P_{\mathrm{s}} + P_{\mathrm{Joule}},\label{eq:temp}
	\end{equation}
	with suitable boundary and initial conditions, where $\rho$ is the mass density, $C_{\mathrm{p}}$ the heat capacity, $k$ the thermal conductivity 
	and $T$ the temperature. The Joule losses $P_{\mathrm{Joule}}$ are defined as
	\begin{equation*}
	P_{\mathrm{Joule}} = q_{\mathrm{flag}}\,\sigma^{-1} \|\vec{J}_{\mathrm{s}}\|^2
	\end{equation*}
	with $\sigma$ being the homogenized non-linear conductivity of the cables in the case of a quench. It is activated by the sigmoid-type activation 
	function $q_{\mathrm{flag}}$ that depends on time, the magnetic flux density $\vec{B}$ and current density $\vec{J}_{\mathrm{s}}$.
	$P_{\mathrm{s}}$ is the power density in the superconductor 
	and relates the heat equation with the magnetoquasistatic field solution through
	\begin{equation*}
	P_{\mathrm{s}} = \vec{M}_{\mathrm{s}}\cdot\frac{\partial\vec{B}}{\partial t}.
	\end{equation*}
	In case of quenching, i.e., $q_{\mathrm{flag}}>0$, the superconducting coils have an Ohmic resistance with the voltage drop
	\begin{equation}
	\mathbf{v}_\text{t} = \mathbf{R}\mathbf{i}
	\text{ with }
	\mathbf{R}=q_{\mathrm{flag}}\int_{\Omega_\text{s}}\vec{\chi}_{\mathrm{s}}^{\,\top}\sigma^{-1}\vec{\chi}_{\mathrm{s}} \ \mathrm{d} \Omega. 
	\label{eq:temp_coup}
	\end{equation}
	The non-linear resistance $\mathbf{R}$ inherits the dependencies on $t$, $\vec{B}=\nabla\times\vec{A}$ and 
	$\vec{J}_\text{s}=\vec{\chi}_{\mathrm{s}}\mathbf{i}$.

	After spatial discretization of (\ref{eq:mqs}-\ref{eq:temp_coup}) as e.g. obtained by the Finite Element (FE) Method \cite{Monk_2003aa}, 
	the semi-discrete system 
	\begin{subequations}
		\label{eq:mqs_thermal}
		\begin{align}
		\mathbf{K}_{\mathrm{s}}\bigl(\mathbf{a}\bigr)\frac{\mathrm{d}\mathbf{a}}{\mathrm{d}t}+\mathbf{K}_{\nu}\bigl(\mathbf{a}\bigr)\mathbf{a} - 
		\mathbf{X}\mathbf{i}&= 0 \label{eq:mqs_fem1}\\
		\mathbf{v} -\mathbf{X}^\top\frac{\mathrm{d}\mathbf{a}}{\mathrm{d}t} &= 0\label{eq:mqs_fem2}\\
		\mathbf{M}_{\rho}\frac{\mathrm{d}\mathbf{t}}{\mathrm{d}t}+\mathbf{K}_{\mathrm{k}}\mathbf{t} - \mathbf{q}(\mathbf{a})&= 
		0\label{eq:thermal1}\\
		\mathbf{v}_\text{t} -\mathbf{R}(\mathbf{t},\mathbf{a},\mathbf{i})\mathbf{i} &= 0\label{eq:thermal2}
		\end{align}
	\end{subequations}
	is obtained, where $\mathbf{a}$ denotes the degrees of freedom for the magnetic vector potentials, $\mathbf{t}$ the discrete temperatures, 
	$\mathbf{q}(\mathbf{a})$ the discretized Joule losses, $\mathbf{M}_\star$ and $\mathbf{K}_\star$ the discretized material and differential 
	operator matrices and $\mathbf{X}$ 
	is the discretization of $\vec{\chi}_\text{s}$.
	Initial conditions $\mathbf{v}_0$, $\mathbf{v}_{\mathrm{t,0}}$, 
	$\mathbf{a}_0$ and $\mathbf{t}_0$ at time $t_0$ are set.
	The system is differential algebraic, as $\mathbf{K}_{\mathrm{s}}$ is a singular matrix due to the fact that $\tau_{\mathrm{eq}}$ is only 
	non-zero in the coil 
	region (and additionally in the 3D case as a consequence of the null space of the curl-curl operator).

	The equations describing the overall behavior
	of the circuit 
	 can be written in an abstract form as the following system
	\begin{subequations}
		\label{eq:cir}
		\begin{align}
			\ensuremath{ \mathbf{A} } \frac{\mathrm{d}\ensuremath{ \mathbf{x} }}{\mathrm{d}t} + \ensuremath{ \mathbf{B} }\bigl(\ensuremath{ 
			\mathbf{x} }\bigr) \ensuremath{ \mathbf{x} } + \ensuremath{ \mathbf{P} } \ensuremath{ \mathbf{i} }&= \mathbf{f}(t)\\
			\ensuremath{ \mathbf{P} }^{\top}\ensuremath{ \mathbf{x} } - \ensuremath{ \mathbf{v} } &= 0\\
			\ensuremath{ \mathbf{Q} }^{\top} \ensuremath{ \mathbf{x} } - \ensuremath{ \mathbf{v} }_\text{t} &= 0.
		\end{align}
	\end{subequations}
	with initial conditions $\mathbf{x}(t_0) = \mathbf{x}_0$ and $\mathbf{i}(t_0) = \mathbf{i}_0$.
	In the case of modified nodal analysis (see \cite{Ho_1975aa}), which is used in most modern circuit simulators, the degrees of freedom 
	$\mathbf{x}$ 
	contain the electric potentials at the circuit nodes and currents through the circuit branches associated with vol\-tage sources, inductors and 
	magnets 
	(coupled via the voltages $\mathbf{v}$ and $\mathbf{v}_\text{t}$).

	\section{Waveform Relaxation}
	 Following \cite{Schops_2010aa}, waveform relaxation is used to simulate the coupled problem (\ref{eq:mqs_thermal}-\ref{eq:cir}). In this 
	 approach, the magnetothermal problem \eqref{eq:mqs_thermal} is simulated separately from the circuit \eqref{eq:cir}. Information between the two 
	 systems is exchanged only at certain synchronization points $\bar{t}_0,\ldots,\bar{t}_n$. Among other advantages, this exploits the different 
	 time scales at which significant effects occur in each subproblem, as the time stepper of each problem can choose its step sizes independently.
	 A Gauss-Seidel-type scheme on a time window $\mathcal{I}_j=[\bar{t}_{j},~\bar{t}_{j+1}]$ using current and voltages as `transmission conditions' 
	 is given by 
	 System 1 (circuit, 
	 cf. \eqref{eq:cir})
	 \begin{subequations}
	 	\label{eq:cir_1}
	 	\begin{align}
	 	\mathbf{A}_\mathrm{c}
	 	\frac{\mathrm{d}}{\mathrm{d}t}\mathbf{x}^{(k+1)} + 
	 	\mathbf{B}_\mathrm{c}
	 	\mathbf{x}^{(k+1)} + \mathbf{P}\mathbf{i}_{\mathrm{c}}^{(k+1)} &= 
	 	{\textbf{f}}(t) \label{eq0.1}\\
	 	\mathbf{v}_{\mathrm{c}}^{(k+1)} &= \mathbf{P}^{\top}\mathbf{x}^{(k+1)} \label{eq0.2}\\
	 	\mathbf{v}^{(k+1)}_{\mathrm{t,c}}&=\mathbf{Q}^{\top}\mathbf{x}^{(k+1)}\label{eq0.25}
	 	\intertext{with the transmission conditions}
	 	\mathbf{v}_{\mathrm{c}}^{(k+1)} &= \mathbf{v}_{\mathrm{m}}^{(k)} \label{eq0.3}\\
	 	\mathbf{v}^{(k+1)}_{\mathrm{t,c}}&=\mathbf{v}_{\mathrm{t,m}}^{(k)}
	 	\end{align}
	 \end{subequations}	 
	 and System~2 (magnetothermal, 
	 cf. \eqref{eq:mqs_thermal})
	 \begin{subequations}
	 	\label{eq:mag_1}
	 	\begin{align}
	 	\mathbf{K}_{\mathrm{s}} \frac{\mathrm{d}}{\mathrm{d}t}\mathbf{a}^{(k+1)} + \mathbf{K}_{\nu} \mathbf{a}^{(k+1)} &= \mathbf{X} 
	 	\mathbf{i}_{\mathrm{m}}^{(k+1)} \label{eq0.4}\\
	 	\mathbf{X}^{\top} \frac{\mathrm{d}}{\mathrm{d}t}\mathbf{a}^{(k+1)} &=\mathbf{v}_{\mathrm{m}}^{(k+1)} \label{eq0.5}\\
	 	\mathbf{M}_{\rho}\frac{\mathrm{d}}{\mathrm{d}t}\mathbf{t}^{(k+1)}+\mathbf{K}_{\mathrm{k}}\mathbf{t}^{(k+1)} 
	 	&=\mathbf{q}\left(\mathbf{a}^{(k+1)}\right)\\
	 	\mathbf{v}_{\mathrm{t,m}}^{(k+1)}&=\mathbf{R}^{(k+1)}\mathbf{i}^{(k+1)}_{\mathrm{m}}\label{eq:res}
	 	\intertext{with the transmission condition}
	 	\mathbf{i}_{\mathrm{m}}^{(k+1)} &= \mathbf{i}_{\mathrm{c}}^{(k+1)}.\label{eq0.6}
	 	\end{align}
	 \end{subequations}
	 The algorithm
	 consists of the following steps\\[-0.5em] 
	 \begin{itemize}
	 	\item[0)] set 
	 	$\mathbf{v}_{\mathrm{m}}^{(0)}(t) = \mathbf{v}_{0}$ and $\mathbf{v}_{\mathrm{t,m}}^{(0)}(t) = 
	 	\mathbf{v}_{\mathrm{t},0}$ on $t\in\mathcal{I}_0$\\[-0.5em]
	 	\item[1)] solve \eqref{eq:cir_1} for given inputs $\mathbf{v}_{\mathrm{m}}^{(k)}(t)$ and $\mathbf{v}_{\mathrm{t,m}}^{(k)}(t)$ on 
	 	$\mathcal{I}_j$ to obtain the output $\mathbf{i}_{\mathrm{c}}^{(k+1)}(t)$ \\[-0.5em]
	 	\item[2)] solve \eqref{eq:mag_1} for given input $\mathbf{i}_{\mathrm{c}}^{(k+1)}(t)$ on $\mathcal{I}_j$ to obtain the outputs 
	 	$\mathbf{v}_{\mathrm{m}}^{(k+1)}(t)$ and $\mathbf{v}_{\mathrm{t,m}}^{(k+1)}(t)$ \\[-0.5em]
	 	\item[3)] if not converged, set $k=k+1$ and go to step 1), otherwise proceed with 
	 	$\mathbf{v}_{\mathrm{m}}^{(0)}(t) = 
	 	\mathbf{v}_{\mathrm{m}}^{(k+1)}(\bar{t}_\mathrm{j})$, $\mathbf{v}_{\mathrm{t,m}}^{(0)}(t) = 
	 	\mathbf{v}_{\mathrm{t,m}}^{(k+1)}(\bar{t}_\mathrm{j})$, for 
	 	$t\in\mathcal{I}_\mathrm{j+1}$, $j=j+1$ 
	 	and go to step 1).\\[-0.5em]
	 \end{itemize}
	 Here, initialization of the voltages at each waveform relaxation window is carried out via constant extrapolation of its value at 
	 the last time step of the previous window. However, this initialization can also be implemented differently, e.g. by using higher order 
	 extrapolation.

	It shall be noted that the excitation of the field problem \eqref{eq:mag_1} by a known voltage source instead of a current source increases the 
	numerical stability, 
	as the differential algebraic index of the system is minimized \cite{Bartel_2011aa}. However, for this paper this aspect is disregarded and the 
	field problem is considered 
	in the circuit as a current-driven voltage source. This corresponds, from a Schwarz-type methods point of view, to using `zero-order' Dirichlet 
	transmission conditions. As 
	in \cite{Al-Khaleel_2014aa} and \cite{Nshimiyimana_2016aa}, the waveform relaxation scheme can be studied from the point of view of an optimized 
	Schwarz method, such that the 
	transmission conditions are optimized to speed up the convergence in terms of $k$.

\subsection{Optimized waveform relaxation}
	The following analysis is carried under several simplifying assumptions: only the coupling between electromagnetic field 
	(\ref{eq0.4}-\ref{eq0.5}) 
	and circuit (\ref{eq0.1}-\ref{eq0.2}) is considered, the thermal case can be treated analogous. Moreover, the reluctivities $\nu$ and 
	$\nu_\text{s}$, 
	as well as the time constant $\tau_{\mathrm{eq}}$ are kept constant. It is shown later that these steps do not limit the applicability in 
	practice. 
	
	Let us rewrite the system in frequency domain using the angular frequency $\omega=2\pi f$. Then
	\begin{subequations}
		\begin{align}
		\mathbf{A}_\mathrm{c}
		j \omega\mathbf{x}^{(k+1)} + 
		\mathbf{B}_\mathrm{c}
		\mathbf{x}^{(k+1)} + \mathbf{P}\mathbf{i}_{\mathrm{c}}^{(k+1)} &= 
		{\textbf{g}}(\omega) \label{eq1}\\
		\mathbf{v}_{\mathrm{c}}^{(k+1)} &= \mathbf{P}^{\top}\mathbf{x}^{(k+1)} \label{eq2}\\
		\mathbf{v}_{\mathrm{c}}^{(k+1)} &= \mathbf{v}_{\mathrm{m}}^{(k)} \label{eq3}
		\end{align}
	\end{subequations}
	and 
	\begin{subequations}
		\begin{align}
		\mathbf{K}_{\mathrm{s}} j\omega\mathbf{a}^{(k+1)} + \mathbf{K}_{\nu} \mathbf{a}^{(k+1)} &= \mathbf{X} \mathbf{i}_{\mathrm{m}}^{(k+1)} 
		\label{eq4}\\
		\mathbf{X}^{\top} j \omega\mathbf{a}^{(k+1)} &=\mathbf{v}_{\mathrm{m}}^{(k+1)} \label{eq5}\\
		\mathbf{i}_{\mathrm{m}}^{(k+1)} &= \mathbf{i}_{\mathrm{c}}^{(k+1)}.\label{eq6}
		\end{align}
	\end{subequations}
	For the following optimization of convergence, the first transmission condition \eqref{eq3} is generalized to the linear combination
	\begin{equation}
	\mathbf{v}_{\mathrm{c}}^{(k+1)} = \alpha\mathbf{i}_{\mathrm{c}}^{(k+1)} - \alpha\mathbf{i}_{\mathrm{m}}^{(k)} + \mathbf{v}_{\mathrm{m}}^{(k)}, 
	\label{eq7} 
	\end{equation}
	while the second transmission condition \eqref{eq6} is kept in its original form since its optimization is cumbersome for arbitrary circuitry. 

The following section discusses how the weighting factor $\alpha$ influences the convergence of the waveform relaxation scheme.

\subsection{Optimization of contraction factor in frequency domain}
Waveform relaxation is a fixed point iteration in time domain and therefore its convergence can only be guaranteed if there is a contraction of the 
error, see e.g. 
\cite{Burrage_1995aa}. This is expressed by a contraction factor $\rho(\alpha)<1$, such that
\begin{equation*}
||\mathbf{v}_{\mathrm{c}}^{(k+1)}- \mathbf{v}_{\mathrm{c}}^{(k)}|| = |\rho(\alpha)| \ 
||\mathbf{v}_{\mathrm{c}}^{(k)}-\mathbf{v}_{\mathrm{c}}^{(k-1)}||.
\end{equation*}
Assuming that
$(\mathbf{K}_{\mathrm{s}}j\omega + \mathbf{K}_{\nu})$ is 
invertible, which is necessary for the solvability of the subsystem, we obtain
\begin{equation*}
\rho(\alpha) = \left(\textbf{I} + \alpha\mathbf{x}_{\mathrm{P}}^{-1}(\omega) \right)^{-1}\big(\alpha - 
\mathbf{Z}(\omega)\big)\mathbf{x}_{\mathrm{P}}^{-1}(\omega),
\end{equation*}
where we assume that
the transfer impedance $\mathbf{x}_{\mathrm{P}}(\omega) = \mathbf{P}^\top(\mathbf{A}_\mathrm{c} j \omega + 
\mathbf{B}_\mathrm{c})^{-1}\mathbf{P}$ exists and is invertible
and 
the impedance of the magnetoquasistatic system 
\begin{align}\label{eq:optimal}
\mathbf{Z}(\omega) = j\omega\mathbf{X}^\top(\mathbf{K}_{\mathrm{s}}j\omega + \mathbf{K}_{\nu})^{-1}\mathbf{X}.
\end{align}
Optimal convergence is attained for $\rho(\alpha) = 0$, that is
\begin{equation*}
\alpha = \mathbf{Z}(\omega).
\end{equation*}
Therefore, the optimized transmission condition for the circuit is given by
\begin{equation*}
\mathbf{v}_{\mathrm{c}}^{(k+1)} = \mathbf{Z}(\omega)\mathbf{i}_{\mathrm{c}}^{(k+1)} - \mathbf{Z}(\omega)\mathbf{i}_{\mathrm{m}}^{(k)} + 
\mathbf{v}_{\mathrm{m}}^{(k)}.
\end{equation*}
In frequency domain, the impedance $\mathbf{Z}$ is computable and no further (numerical) optimization is needed to obtain an optimal transmission 
condition. However, when 
proceeding to non-linear systems in time domain, $\mathbf{Z}$ contains time derivatives which need to be approximated. 

\subsection{Approximation of the optimal condition by inductances}
	Let us assume that $\mathbf{K}_{\nu}$ is invertible. Then, we can rewrite the impedance as
	\begin{align*}
	\mathbf{Z}(\omega)&=j\omega\mathbf{X}^\top\mathbf{K}_{\nu}^{-\frac{1}{2}}\left(\mathbf{I} + 
	j\omega\mathbf{K}_{\nu}^{-\frac{1}{2}}\mathbf{K}_{\mathrm{s}}\mathbf{K}_{\nu}^{-\frac{1}{2}}\right)^{-1}\mathbf{K}_{\nu}^{-\frac{1}{2}}\mathbf{X},
	\intertext{which can be recast in a Neumann series}
	\mathbf{Z}(\omega)&=j\omega\mathbf{X}^\top\mathbf{K}_{\nu}^{-\frac{1}{2}}\sum_{l=0}^{\infty}\left(-j\omega\mathbf{K}_{\nu}^{-\frac{1}{2}}\mathbf{K}_{\mathrm{s}}\mathbf{K}_{\nu}^{-\frac{1}{2}}\right)^l\mathbf{K}_{\nu}^{-\frac{1}{2}}\mathbf{X},\label{eq:series}
	\end{align*}
	if the operating frequency $\omega$ is low, as it must hold
	\begin{equation*}
	\kappa:=|| -j\omega\mathbf{K}_{\nu}^{-\frac{1}{2}}\mathbf{K}_{\mathrm{s}}\mathbf{K}_{\nu}^{-\frac{1}{2}}||<1.
	\end{equation*}
	To analyze the smallness of $\kappa$, we make some further assumptions:
	let the magnetoquasistatic system be discretized, such that the curl-curl operators can be factorized as 
	$\mathbf{K}_{\star}=\mathbf{C}^{\top}\mathbf{M}_{\star}\mathbf{C}$ with diagonal material matrices $\mathbf{M}_{\star}$. Then, it can be shown 
	that 
	$$f_{\mathrm{max}} < \frac{1}{2\pi\tau_{\mathrm{max}}}$$
	holds, since 
	\begin{align*}
	\kappa\leq\omega_{\mathrm{max}}||\mathbf{K}_{\nu}^{-\frac{1}{2}}\mathbf{C}^\top 
	\mathbf{M}_{\tau_{\mathrm{max}}}\mathbf{M}_{\nu}\mathbf{C}\mathbf{K}_{\nu}^{-\frac{1}{2}}||,
	\end{align*}
	with $\mathbf{M}_{\tau_{\mathrm{max}}} =\max\limits_i(\mathbf{M}_{\tau})_{ii}\mathbf{I} = \mathbf{M}_{\tau} + \bar{\mathbf{M}}_\tau$, and 
	$\bar{\mathbf{M}}_\tau$ a positive semidefinite diagonal matrix. $\mathbf{M}_{\nu} = \mathbf{M}_{\nu_{\mathrm{s}}} + 
	\bar{\mathbf{M}}_{\nu_{\mathrm{s}}}$, with $\bar{\mathbf{M}}_{\nu_{\mathrm{s}}}$ also positive semidefinite, as 
	$(\mathbf{M}_{\nu_{\mathrm{s}}})_{ii}=(\mathbf{M}_{\nu})_{ii}$ in the coil domain and zero everywhere else. Therefore the series converges if
	$$\omega_{\mathrm{max}}\tau_{\mathrm{max}} < 1.$$
	Finally, the optimal coefficient \eqref{eq:optimal} can be approximated by a finite number of terms of the Neumann series, e.g. for $l=0$, it 
	follows that
	\begin{equation}\label{eq:induc2}
	\mathbf{Z}(\omega) \approx j\omega \mathbf{L} := j\omega\mathbf{X}^\top\mathbf{K}_{\nu}^{-1}\mathbf{X}.
	\end{equation}
	Using the coefficient $\alpha=j\omega\mathbf{L}$ and dividing the transmission condition \eqref{eq7} by $j\omega$ yields 
	\begin{equation}\label{eq:induc}
	\flux_{\mathrm{c}}^{(k+1)} = \mathbf{L} \mathbf{i}_{\mathrm{c}}^{(k+1)} - \mathbf{L} \mathbf{i}_{\mathrm{m}}^{(k)} + \flux_{\mathrm{m}}^{(k)},
	\end{equation}
	with the magnetic flux linkage $\flux_\star=\frac{1}{j\omega}\mathbf{v}_\star$. The inductance can be obtained by solving the magnetostatic 
	problem, 
	i.e. \eqref{eq:mqs}, with $\tau=0$ excited by $1$\,A in each coil. This extraction comes with moderate additional numerical costs, since only one 
	(linear) system of equations has to be solved with several right hand sides. The transmission condition \eqref{eq:induc} corresponds to 
	considering 
	the field model in the circuit as an inductance with a correction term as proposed in \cite{Schops_2011ac}. Considering further terms of the 
	Neumann 
	series is possible but has not been investigated. This would require higher regularity of the solution since higher order time derivatives are 
	needed.

Although the analysis above has been carried out for a li\-ne\-a\-ri\-za\-tion of the original system and under restrictive assumptions on the 
discretization, in practice, 
it has been observed that the convergence of the waveform relaxation scheme with the transmission condition \eqref{eq:induc} is substantially 
improved. The following section 
discusses the non-linear case.

\subsection{Improved algorithm in time domain}\label{sec:algo}
When treating non-linear field/circuit coupled problems with waveform relaxation, the inductance in the transmission condition \eqref{eq:induc} is 
time dependent due to the magnetic saturation in \eqref{eq:mqs_thermal}, i.e. 
$$\mathbf{L}^{(k)}(t) := \mathbf{X}^\top\mathbf{K}_{\nu}^{-1}\left(\mathbf{a}^{(k)}(t)\right)\mathbf{X}.$$ 
Following this approach rigorously requires to continuously update the inductance, e.g. by repeating the inductance extraction described in the 
previous section after each successful time step. In contrast, we propose a simplified procedure: the differential inductance 
\begin{equation}\label{eq:linearized_Lm}
\mathbf{L}_{\text{m}} = \mathbf{X}^\top
\mathbf{K}_{\nu,\mathrm{d}}^{-1}(\mathbf{a}_{\star})
\mathbf{X}
\end{equation}
is extracted at a working point $\mathbf{a}_{\star}$ from the differential curl-curl matrix 
$\mathbf{K}_{\nu,\mathrm{d}}=\frac{\mathrm{d}}{\mathrm{d}\mathbf{a}}\Bigl(\mathbf{K}_{\nu}\bigl(\mathbf{a}\bigr)\mathbf{a}\Bigr)$
and it is kept constant for the subsequent waveform relaxation, i.e., the condition \eqref{eq3} is replaced by
\begin{equation}\label{eq:vdiff}
\mathbf{v}_\mathrm{c}^{(k+1)}(t)=\mathbf{L}_{\text{m}}\frac{\mathrm{d}}{\mathrm{d}t}\mathbf{i}_{\mathrm{c}}^{(k+1)}(t) + 
\Delta\mathbf{v}_{\text{m}}^{(k)}(t),
\end{equation}
which corrects the misfit due to disregarding the non-linearity and the higher order terms of the Neumann series by a lumped voltage source 
\begin{equation}\label{eq:voltageDiff}
\Delta\mathbf{v}_{\text{m}}^{(k)}(t):=\frac{\mathrm{d}}{\mathrm{d}t}\flux_{\mathrm{m}}^{(k)}(t)- 
\mathbf{L}_{\text{m}}\frac{\mathrm{d}}{\mathrm{d}t}\mathbf{i}_{\mathrm{m}}^{(k)}(t).
\end{equation}

Inspired by the inductance coupling in the field-circuit case, resistances are now passed from the thermal equation to the circuit instead of the 
voltages in equation \ref{eq:res} through
\begin{equation}\label{eq:res_coupl}
\mathbf{v}_\mathrm{t,c}^{(k+1)} = \mathbf{R}^{(k)}(t)\mathbf{i}_\mathrm{c}^{k+1},
\end{equation}
where $\mathbf{R}^{(k)}(t) = \mathbf{R}(t, \mathbf{a}^{(k)}, \mathbf{i}_\mathrm{m}^{(k)})$. 
The Gauss-Seidel-type scheme using these new transmission conditions reads\\[-0.5em]
\begin{itemize}
	\item[0)] set $k=0$, 
	$\Delta\mathbf{v}_{\mathrm{m}}^{(0)}(t) = \Delta\mathbf{v}_{\mathrm{m, 0}}(t)$ and \hphantom{and R =R()}
	$\mathbf{R}^{(0)}(t) = 
	\mathbf{R}(t_0,\mathbf{a}_0, \mathbf{i}_0)$ and extract $\mathbf{L}_{\text{m}}$ for $\mathbf{a}_{\star}=\mathbf{a}_0$ using 
	\eqref{eq:linearized_Lm} 
	\\[-0.5em]
	\item[1)] solve (\ref{eq0.1}-\ref{eq0.25}) with conditions \eqref{eq:vdiff} and \eqref{eq:res_coupl} for given inputs 
	$\Delta\mathbf{v}_{\text{m}}^{(k)}(t)$ and $\mathbf{R}^{(k)}(t)$ to obtain the 
	output 
	$\mathbf{i}_{\mathrm{c}}^{(k+1)}(t)$\\[-0.5em]
	\item[2)] solve \eqref{eq:mag_1} for given input $\mathbf{i}_{\mathrm{c}}^{(k+1)}(t)$ to obtain the output voltage 
	$\Delta\mathbf{v}_{\text{m}}^{(k)}(t)$ according to \eqref{eq:vdiff} and $\mathbf{R}^{(k+1)}(t)$ as before\\[-0.5em]
	\item[3)] if not converged, set $k=k+1$ and go to step 1), otherwise proceed with 
	$\Delta\mathbf{v}_{\mathrm{m}}^{(0)}(t) = 
	\Delta\mathbf{v}_{\mathrm{m}}^{(k+1)}(\bar{t}_\mathrm{j})$, $\mathbf{R}^{(0)}(t) = 
	\mathbf{R}^{(k+1)}(\bar{t}_\mathrm{j})$, $j=j+1$ 
	and go to step 1).\\[-0.5em]
\end{itemize}
Obviously, the algorithm can be improved by appropriate heuristics to update $\mathbf{L}_{\text{m}}$ if the number of iterations $k$ grows.

\begin{figure}	
	\centering
	\includegraphics[scale = 1]{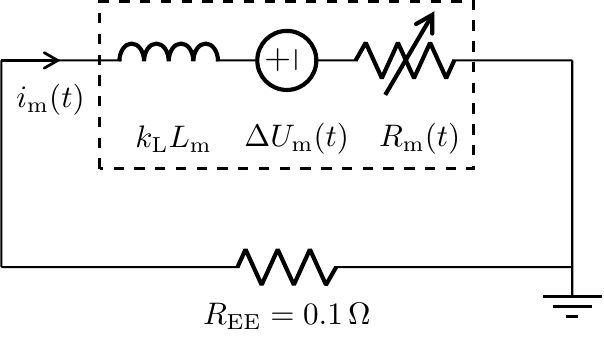}
	\caption{Schematic of the protection circuit with a lumped model representing the magnet (dashed box),
		where $R_{\mathrm{m}}(t)$ 
		is defined in equation \eqref{eq:resistance}, $\Delta v_{\mathrm{m}}(t)$ in \eqref{eq:voltageDiff} and $L_{\mathrm{m}}$ in 
		\eqref{eq:inductance}. Finally, $i_\mathrm{m}(t)$ is the current through the magnet.}
	\label{fig:CircuitSchematic}
\end{figure}

\section{Numerical Examples}
Let us first discuss the convergence of the Neumann series in practice and then give an example for the field/circuit quench simulation of the D1 
accelerator magnet \cite{Nakamoto_2015aa}. 
\subsection{Discussion of the time constant}\label{sec:time_const}
The time constant $\tau_{\mathrm{eq}}$ depends on the cable parameters and is mathematically expressed by \cite{Verweij_1995aa}
\begin{equation*}
\tau_{\mathrm{eq}}(B) = \frac{\mu_0}{2}\left(\frac{l_{\mathrm{f}}}{2\pi}\right)^2\frac{1}{(c_0+c_1B)f_{\mathrm{eff,x}}},
\end{equation*}
with $l_{\mathrm{f}}$ being the filament twist-pitch, $\mu_0$ the vacuum permeability, $f_{\mathrm{eff,x}}$ the fraction of superconductor in the 
matrix, $c_0$ and $c_1$ 
being characterized by the resistivity of the material used in the matrix and $0< B\leq 10$ being the magnitude of the magnetic flux density.

Typical values for those parameters are given in \cite{Ravaioli_2015aa}. For the case of a Nb-Ti dipole accelerator magnet, one has
\begin{align*}
	&l_{\mathrm{f}} = 1.5 \cdot10^{-2}\,\mathrm{m}, && c_0 = 1.7 \cdot 10^{-10}\,\Omega\mathrm{m}\\
	&c_1 = 4.2 \cdot 10^{-11}\,\Omega\mathrm{m}\mathrm{T}^{-1}, && f_{\mathrm{eff,x}} = 1
\end{align*}
and thus it follows
\begin{align*}
	\tau_{\mathrm{eq}} < \frac{\mu_0}{2}\left(\frac{l_{\mathrm{f}}}{2\pi}\right)^2\frac{1}{c_0f_{\mathrm{eff,x}}} < 0.0211\,\mathrm{s},
\end{align*}
which yields convergence of the Neumann series for frequencies below $f_{\mathrm{max}} < 7.5$\,Hz.

\begin{figure}
	\centering
	\includegraphics[scale = 1.1]{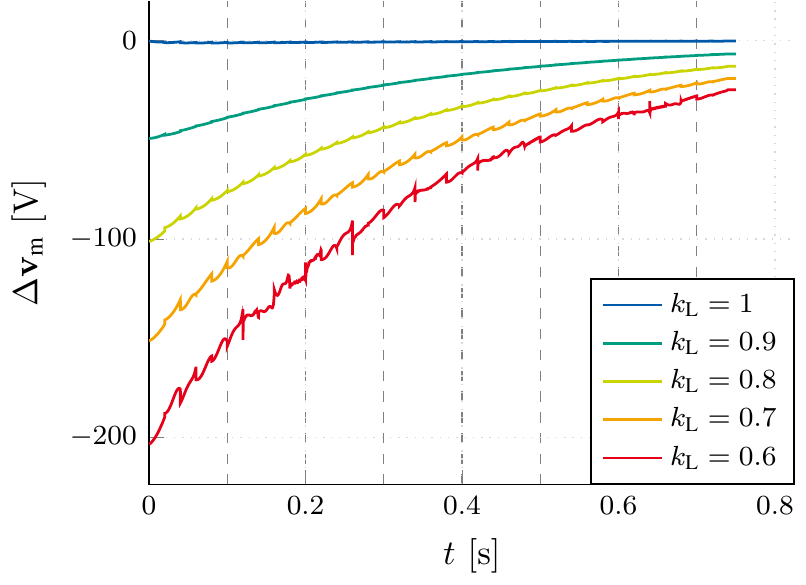}
	\caption{Voltage differences $\Delta\mathbf{v}_{\text{m}}$ correcting the mismatch between field and circuit model in the magnetostatic case}
	\label{fig:VoltageDifferenceMS}
\end{figure}

\subsection{Simulation of dipole magnet}
The described method is illustrated for a numerical example of the single aperture dipole magnet D1 \cite{Nakamoto_2015aa}. The magnet is designed to 
be used for beam separation for the High Luminosity upgrade of the Large Hadron Collider. The dipole magnet is represented by the strongly coupled 
magnetothermal system~\eqref{eq:mqs_thermal} in a 2D setting. Only one quarter is modeled with homogeneous Dirichlet boundary conditions at the outer 
boundaries and Neumann conditions at the symmetry planes
\begin{align*}
\vec{A}|_{\Gamma_{\text{dir}}}\times\vec{n}=0\,\text{Wb/m},
\quad
\bigl(\nu\nabla\times\vec{A}\bigr)\bigl|_{\Gamma_{\text{neu}}}\times\vec{n}=0\,\text{A/m}
\end{align*}
and $k\nabla T|_{\Gamma_\text{s}}\cdot\vec{n}=0$\,W/m$^2$, where $\vec{n}$ is the outward normal vector. The partial differential equations 
\eqref{eq:mqs_fem1}-\eqref{eq:thermal2} are discretized and solved 
using \textsc{COMSOL Multiphysics\textsuperscript{\textregistered}} \cite{COMSOL_2016aa}. The model consists of first 
order and second order nodal elements for the 
thermal and the magnetic problem, respectively, adding up to 9871 degrees of freedom in total.
The semi-discrete problem is time-discretized by a backward differentiation formula (variable order, maximum time step size $\delta t=1$\,ms). All 
parameters, e.g. non-linear materials laws for the reluctivity $\nu$ and time constant $\tau$, are defined as specified in \cite{Nakamoto_2015aa, 
	Bruning_2004aa}. We consider a magnet operating at $5$\,kA and protected by a resistor  $R_{\text{EE}}=0.1\,\Omega$ as depicted in 
Fig.~\ref{fig:CircuitSchematic}. The circuit is modeled and simulated in \textsc{OrCAD PSpice\textsuperscript{\textregistered}} using the 
trapezoidal method (maximum time step $\delta t=10\,\mu$s) 
\cite{OrCAD_2016aa}. The waveform relaxation is carried out on $\mathcal{I}=(0\,\mathrm{s}, 0.75\,\mathrm{s}]$ with initial values $\vec{A}_0$, 
$T_0$  
and $\mathbf{x}_0$ that correspond to a system ramped up to $5$\,kA. The ramped-up values are obtained by initializing the circuit and magnetothermal 
system 
with zero conditions and then solving both problems independently, with current sources that are linearly increased from $0$ to $5$\,kA. The 
co-simulation was established within CERN's in-house coupling tool STEAM (Simulation of Transient Effects in Accelerator Magnets) 
\cite{Bortot_2016ab}.

\begin{figure}
	\centering
	\includegraphics[scale = 1.1]{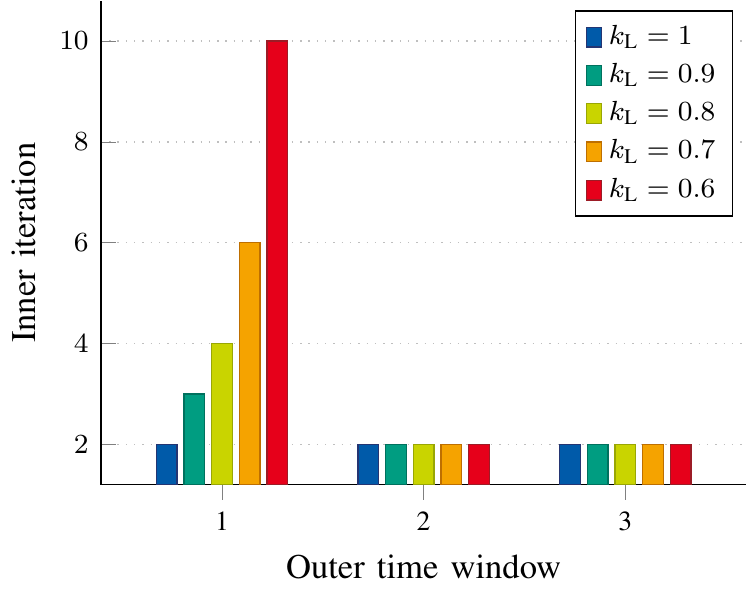}
	\caption{Convergence comparison for first 3 time windows of the magnetostatic case}
	\label{fig:ConvergenceComparisonMS}
\end{figure}

The aim of this study is to analyze the influence of the differential inductance estimation \eqref{eq:linearized_Lm} on the waveform relaxation 
convergence. For that purpose, we introduce a scaling coefficient $k_L$ to be multiplied by the (scalar) differential inductance 
\begin{equation}\label{eq:inductance}
L_{\text{m}} = \mathbf{X}^\top
\mathbf{K}_{\nu,\mathrm{d}}^{-1}(\mathbf{a}_{0})
\mathbf{X},
\end{equation}
calculated once for the initial value $\mathbf{a}_0 = \mathbf{a}(t_0)$. The loop voltage reads
\begin{align}
i_{\text{m}}(t)\bigl(R_{\text{EE}}+R_{\text{m}}(t)\bigr)+\Delta v_{\text{m}}(t)+k_{\text{L}}L_\text{m}\partial_{t} i_{\text{m}}(t)&=0,
\label{LoopVoltageCircuit}
\end{align}
where $i_{\text{m}}(t)$ is the current flowing through the magnet, 
\begin{align}\label{eq:resistance}
R_{\text{m}} = q_{\mathrm{flag}}\,\mathbf{X}^\top \mathbf{M}_\sigma^{-1}\mathbf{X} 
\end{align}
is the time-dependent coil resistance appearing after a quench and $\Delta v_{\text{m}}(t)$ is the (scalar) voltage source accounting for differences 
between the induced voltage of a FE model and the voltage across the magnet's differential inductance according to \eqref{eq:vdiff}. In the case of 
the dipole, the voltage source is determined by 
\begin{equation*}
\frac{\mathrm{d}}{\mathrm{d}t}\psi(t)=
k_{\text{L}}L_{\text{m}}{\frac{\mathrm{d}}{\mathrm{d}t} }i_{\text{m}}(t) + \Delta v_{\text{m}}(t).
\end{equation*}
This is represented by a time-dependent voltage source in the circuit that linearly interpolates time-discrete data obtained from the field model. 

We will study two simulation scenarios for the magnetostatic and the magnetoquasistatic case with varying $$k_{\text{L}}=[1, 0.9, 0.8, 0.7, 0.6, 
0.5].$$ 
This reduces the inductance extracted for $\mathbf{a}_\star=\mathbf{a}_0$ and can be interpreted as an increase of magnetic saturation with respect 
to the situation optimized for $t_0$.

Both scenarios account for the non-linear iron yoke characteristic, whereas only the latter incorporates the inter-filament coupling currents, i.e., 
non-vanishing $\tau_{\mathrm{eq}}$. The coupling is performed by means of the improved waveform relaxation scheme described in 
Section~\ref{sec:algo}. The time interval is split into $38$ windows $\mathcal{I}_j=[\bar{t}_j, \bar{t}_{j+1}]$. Each window has fixed length 
$H_j=\bar{t}_{j+1}-\bar{t}_j=20\mathrm{ms}$ and communication between solver occurs at time stamps $\bar{t}_j$. Iterations $k$ of each window have 
been carried out until two subsequent currents are below a threshold
\begin{equation*}
\frac{\int_{\bar{t}_j}^{\bar{t}_{j+1}}|i_{\text{m}}^{(k)}(t)-i_{\text{m}}^{(k-1)}(t)|\mathrm{d}t}
{\int_{\bar{t}_j}^{\bar{t}_{j+1}}|i_{\text{m}}^{(k)}(t)|\mathrm{d}t}\leq10^{-3}
\end{equation*}

\begin{figure}
	\centering
	\includegraphics[scale = 1.1]{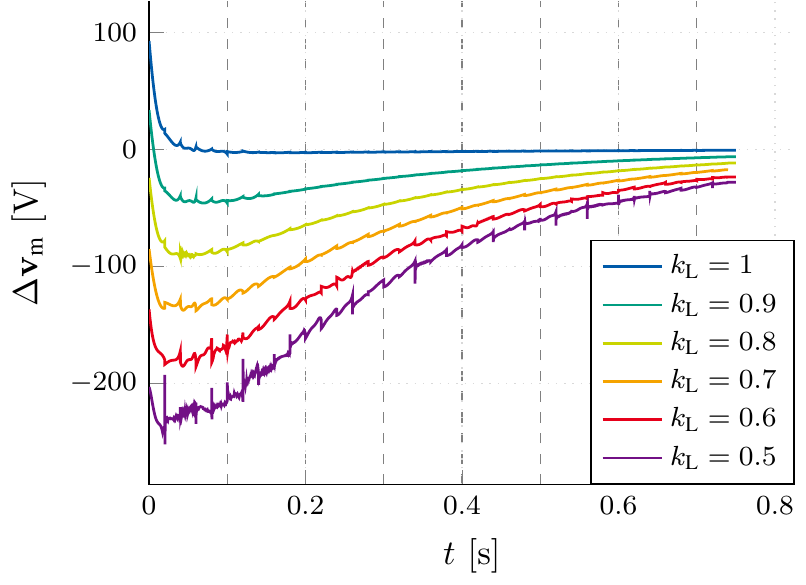}
	\caption{Voltage differences $\Delta\mathbf{v}_{\text{m}}$ correcting the mismatch between field and circuit model in the magnetoquasistatic case}
	\label{fig:VoltageDifferenceMQS}
\end{figure}

\subsection{Magnetostatic model}
The first simulation scenario considers the magnetostatic case, i.e., $\tau_{\mathrm{eq}}\equiv0$ in the partial differential equation 
\eqref{eq:mqs}. 
This simplified case disregards eddy currents and the iron yoke is not saturated at the considered current level such that the magnetic material 
behaves linearly. 
These simple models have been traditionally considered as a first approximation in the simulation of quench protection systems \cite{Rossi_2004aa}. 
The simplification 
facilitates the extraction of an equivalent inductance $L_{\text{m}}$ according to \eqref{eq:linearized_Lm}, which perfectly describes the behavior 
of the partial differential model in the circuit.

Fig.~\ref{fig:VoltageDifferenceMS} summarizes five simulations with varying $k_{\text{L}}=[1,0.9,0.8,0.7,0.6]$ and the given communication time 
window size of $20$\,ms. 
The reference case $k_{\text{L}}=1$ is characterized by a voltage source of $\Delta v_{\text{m}}=0$\,V. This verifies numerically that 
$\alpha=(j\omega L_\text{m})^{-1}$ leads to 
optimal convergence for the magnetostatic model, i.e., it corresponds to an optimal contraction factor of $\rho(\alpha)=0$. For the remaining cases, 
the decrease of $k_{\text{L}}$ is 
compensated by an increased magnitude of $\Delta v_{\text{m}}(t)$. Additional effort needed to obtain convergence of the voltage generator is 
reflected in the increase of number of iterations 
as shown in Fig.~\ref{fig:ConvergenceComparisonMS}. The case $k_{\text{L}}=0.5$ is not depicted, as the iterative scheme diverged.

\begin{figure}
	\centering
	\includegraphics[scale = 1.1]{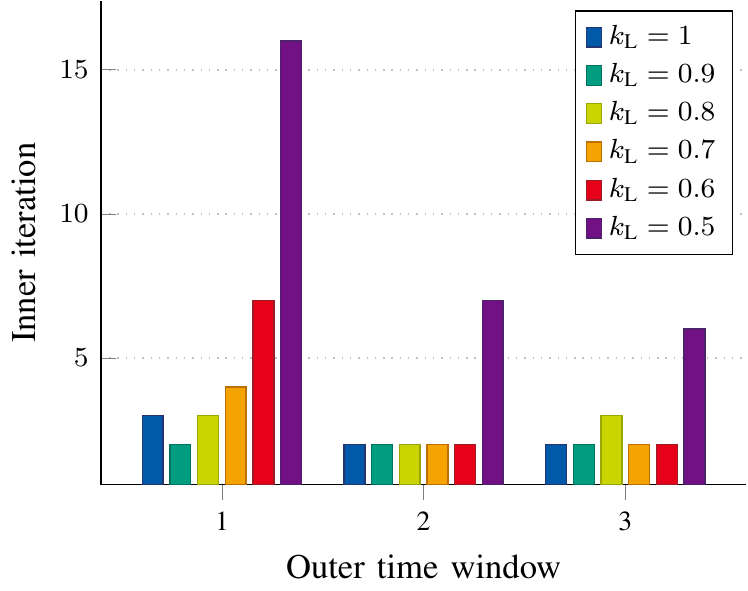}
	\caption{Convergence comparison for first 3 time windows of the magnetoquasistatic case}
	\label{ConvergenceComparisonMQS}
\end{figure}

\subsection{Magnetoquasistatic model}
The second simulation scenario considers the thermal/magnetoquasistatic case using a non-linear time constant $\tau_{\mathrm{eq}}(B)$ as described in 
Section~\ref{sec:time_const}. 
The inter-filament coupling losses are effectively decreasing the flux linkage and as a consequence the induced voltage in the field model.

Fig.~\ref{fig:VoltageDifferenceMQS} shows the simulated voltage differences. For the nominal inductance, i.e. $k_{\text{L}}=1$, the voltage source 
$\Delta v_{\text{m}}(t)$ is initially 
positive and decreases to zero in the consecutive time windows. Fig.~\ref{ConvergenceComparisonMQS} shows the number of iterations needed to obtain a 
sufficiently accurate solution. It is 
worth noticing that for $k_{\text{L}}=0.9$ the convergence is obtained in the least number of iterations as compared to other cases. The reason for 
such convergence lays in inter-filament 
coupling losses, which dissipate the magnetic energy in the coil. These processes results can be attributed to the decrease of the effective 
differential inductance and explains the need to 
inject missing energy by means of $\Delta v_{\text{m}}(t)$. On the other hand, the case of $k_{\text{L}}=0.5$ still converges and leads to correct 
results but requires up to $k=16$ iterations 
instead of the $k=2$ ones for the optimal case.

\section{Conclusion}
This paper has discussed multi-physical field/circuit waveform relaxation for the specific eddy-current model used in quench simulation. In contrast 
to previous works, no numerical 
optimization of the transmission conditions was carried out to obtain optimal convergence rates. Instead, it has been proven for low frequencies that 
the contraction factor is significantly 
reduced by considering an equivalent inductance or higher-order terms of a Neumann series. Numerical simulations of an aperture dipole magnet 
underline the importance of an optimization of the 
coupling conditions, as iterations could be reduced from 16 to 2 per window by using an inductive coupling.

Future research will investigate improved circuit models for loss prediction.

\section*{Acknowledgment}
\noindent This work has been supported by the Excellence Initiative of the German Federal and State Governments and the Graduate School of CE at TU Darmstadt.


\end{document}